\documentclass[12pt]{iopart}
\usepackage{mathptmx}
\usepackage[dvips]{graphics,color,epsfig}
\usepackage{graphicx}

\newcommand{\dd}{\mbox{{\rm d}}}
\newcommand{\GF}{G_{\rm F}}
\newcommand{\diag}{{\rm diag}}

\begin{document}

\title{Three-Neutrino MSW Effect and the Lehmann Mass Matrix}
\author{P Osland}
\address{Department of Physics, University of Bergen, \\
      All\'{e}gaten 55, N-5007 Bergen, Norway}
\begin{abstract}
Recent work on analytical solutions to the MSW equations for three neutrino
flavours is reviewed, with emphasis on the exponential density.  Application
to a particular mass matrix, proposed by Lehmann, Newton and Wu, is also
discussed.  Within this model, the experimental data allow a determination of
the three neutrino masses.  They are found to be
0.002--0.004, 0.01 and 0.05~eV.
\end{abstract}
\section{Introduction}
We here review some recent results on analytical solutions of the
Mikheyev--Smirnov--Wolfenstein (MSW) effect \cite{Wolfenstein:1977ue} for the
propagation of three neutrino flavours.  Such analytic results have been
obtained for both the exponential density
\cite{TorrenteLujan:1995qi,Osland:2000et} and the linear density
\cite{Lehmann:2000ey}.

In the case of an exponential electron density, the three neutrino wave
functions can be expressed in terms of generalized hypergeometric functions,
${}_2F_2$ \cite{TorrenteLujan:1995qi,Osland:2000et}.  For the linear density,
the solutions can be expressed in terms of a Fourier transform
\cite{Lehmann:2000ey}.  In the case of two flavours, these reduce to
parabolic cylinder functions or confluent hypergeometric functions, but of
different parameters and arguments \cite{two-nu,Toshev:jw}.

A particular neutrino mass matrix, originally proposed by Lehmann, Newton and
Wu (LNW) for quarks \cite{Lehmann:1996br} is also reviewed
\cite{Osland:2000bh,Osland:2001kp,Gastmans:2002wm}.  Within this model, the
current data on atmospheric and solar neutrinos permit a determination of the
neutrino masses.
\section{Exponential density}
In a medium where the electron neutrino (described by $\phi_1(r)$)
interacts differently from the others, the propagation is given by the
equation
\begin{equation}
\label{Eq:Schr-1}
i\frac{\dd}{\dd r}
\left[\begin{array}{c}
\phi_1 \\ \phi_2 \\ \phi_3
\end{array} \right]
=\left(\left[
\begin{array}{ccc}
D(r) & 0 & 0 \\
0 & 0 & 0 \\
0 & 0 & 0 \\
\end{array}\right] 
+\frac{1}{2p}
\left[\begin{array}{ccc}
M^2_{11} & M^2_{12} & M^2_{13} \\
M^2_{21} & M^2_{22} & M^2_{23} \\
M^2_{31} & M^2_{32} & M^2_{33} 
\end{array}\right]
\right)
\left[\begin{array}{c}
\phi_1 \\ \phi_2 \\ \phi_3
\end{array}\right]
\end{equation}
The mass matrix is real and symmetric, 
$M^2_{ji}=M^2_{ij}\equiv (M^2)_{ij}$, but otherwise arbitrary.
(The LNW mass matrix will be discussed later.)
Furthermore, 
$D(r)=\sqrt{2}\GF N_e(r)$, with $\GF$ the Fermi constant and
$N_e(r)$ the solar electron density.

For the sun, the density \cite{Bahcall:2000nu}
is well approximated by an exponential,
\begin{equation}
N_e(r) = N_e(0)\, e^{-r/r_0},\quad
r_0\simeq 0.1\times R_\odot .
\end{equation}

Introducing a new radial variable:
$u=r/r_0+u_0$, and performing a rotation on the second and third components,
Eq.~(\ref{Eq:Schr-1}) can be written as
\begin{equation}
\label{Eq:Schr-4}
i\frac{\dd}{\dd u} 
\left[\begin{array}{c}
\psi_1(u) \\ \psi_2(u) \\ \psi_3(u)
\end{array}\right]
=
\left[\begin{array}{ccc}
\omega_1+e^{-u} & \chi_2 & \chi_3 \\
\chi_2 & \omega_2 & 0 \\
\chi_3 & 0 & \omega_3
\end{array}\right] \!
\left[\begin{array}{c}
\psi_1(u) \\ \psi_2(u) \\ \psi_3(u)
\end{array}\right].
\end{equation}

The eigenvalues of the $3\times3$ matrix
\begin{equation}
\left[\begin{array}{ccc}
\omega_1 & \chi_2 & \chi_3 \\
\chi_2 & \omega_2 & 0 \\
\chi_3 & 0 & \omega_3
\end{array}\right] 
\end{equation}
are denoted $\mu_1$, $\mu_2$ and $\mu_3$, 
they are the squares of the neutrino masses multiplied by
$r_0/(2p)$.
Together with $\omega_1$ and $\omega_2$ they control the evolution
of the $\psi_i$.

Introducing now the variable
$z=i e^{-u}$,
the solutions to Eq.~(\ref{Eq:Schr-4}) can be expressed in terms of
solutions to the third-order ordinary differential equation
\begin{eqnarray}
\label{Eq:Schr-6}
&&\biggl[
\left(z\frac{\dd}{\dd z}-i\mu_1\right)
\left(z\frac{\dd}{\dd z}-i\mu_2\right)
\left(z\frac{\dd}{\dd z}-i\mu_3\right)  \nonumber \\
&&\quad
-z\left(z\frac{\dd}{\dd z}-i\omega_2\right)
  \left(z\frac{\dd}{\dd z}-i\omega_3\right)
\biggr]\psi=0,
\end{eqnarray}
namely generalized hypergeometric functions \cite{Osland:2000et,Bateman-1}:
\begin{eqnarray}
&&\psi^{(1)}
=e^{-i\mu_1u} 
 {}_2F_2\biggl[
\begin{array}{cc}
-i(\omega_2-\mu_1), & -i(\omega_3-\mu_1)\\
1-i(\mu_2-\mu_1),    & 1-i(\mu_3-\mu_1)
\end{array}
\bigg|ie^{-u} \biggr] \nonumber \\
&&\psi^{(2)}
=e^{-i\mu_2u} 
 {}_2F_2\biggl[
\begin{array}{cc}
-i(\omega_2-\mu_2), & -i(\omega_3-\mu_2)\\
1-i(\mu_1-\mu_2),    & 1-i(\mu_3-\mu_2)
\end{array}
\bigg|ie^{-u} \biggr] \nonumber \\
&&\psi^{(3)}
=e^{-i\mu_3u} 
 {}_2F_2\biggl[
\begin{array}{cc}
-i(\omega_2-\mu_3), & -i(\omega_3-\mu_3)\\
1-i(\mu_1-\mu_3),    & 1-i(\mu_2-\mu_3)
\end{array}
\bigg|ie^{-u} \biggr] 
\end{eqnarray}
The solutions to Eq.~(\ref{Eq:Schr-4}) are thus
\begin{equation}
\label{Eq:psi_i}
\psi_i=C_1\psi_i^{(1)}+C_2\psi_i^{(2)}+C_3\psi_i^{(3)},
\end{equation}
where the constants $C_j$ are determined by the boundary conditions:
$\psi_1=1$, $\psi_2=\psi_3=0$ at the center of the sun, $r=0$.

The ${}_2F_2$ functions can be defined in terms of the series expansions
\begin{equation}
\label{Eq:F22-series}
{}_2F_2\biggl[
\begin{array}{cc}
\alpha_1, & \alpha_2\\
\rho_1,   & \rho_2
\end{array}
\bigg|z \biggr]
=\sum_{k=0}^\infty
\frac{(\alpha_1)_k(\alpha_2)_k}{(\rho_1)_k(\rho_2)_k}\,
\frac{z^k}{k!} 
\end{equation}
where $(\alpha)_k$ is a Pochhammer symbol,
but (somewhat complicated) asymptotic approximations are actually
more useful for numerical work 
\cite{Osland:2000et,Osland:2000bh,Osland:2001kp}.
\section{The LNW mass matrix}

For quarks, it was found \cite{Lehmann:1996br}
that a particular, simple texture
for the $d$ ($d$, $s$, $b$) and $u$ ($u$, $c$, $t$) quark mass
matrices leads to a good description of the CKM matrix \cite{CKM}.
The mass matrix is assumed to have the form
\begin{equation} \label{Eq:mass-texture}
M=\left[\begin{array}{ccc}
0  & d & 0 \\
d  & c & b \\
0  & b & a
\end{array}\right]
\end{equation}
with $b^2=8c^2$.  This is known as a two-zero texture, but differs from those
normally considered (see, e.g., \cite{Altarelli:gu,Roberts:2001zy}) in the
additional relation between $b$ and $c$.  The eigenvalues are given by $m_1$,
$m_2$, and $m_3$, with $m_1\le m_3$.

Actually, in the case of quarks, since there is CP violation,
a complex CKM matrix is obtained by replacing the parameter $d$
in (\ref{Eq:mass-texture}) by $\pm id$ for $u$ quarks 
(such that $M$ remains Hermitian).
The CKM matrix becomes
\begin{equation}
V_{\rm CKM}=R^{\rm T}(u){\rm diag}(-i,1,1) R(d),
\end{equation}
The Jarlskog determinant \cite{Jarlskog:1985ht}
thus obtained is $J=2.6\times10^{-5}$,
in good agreement with data.

This same mass matrix has been applied to the case of three neutrinos
\cite{Osland:2000bh,Osland:2001kp,Gastmans:2002wm}, and rather good fits to the
atmospheric \cite{Fukuda:1998tw} and solar
\cite{Fukuda:1998fd,Fukuda:1998ua,Cleveland:1994er,Abdurashitov:1999bv,
Hampel:1998xg,Fukuda:2001nj,Ahmad:2001an,Ahmad:2002jz}
neutrino data have been obtained.

The matrix $M$ is diagonalized,
whereby $M=RM_{\rm diag}R^{\rm T}$. For this purpose,
the following notation is convenient
\begin{eqnarray}
S_1&\equiv& m_3-m_2+m_1, \nonumber \\
&=&a+c \nonumber \\
-S_2&\equiv&m_3m_2-m_3m_1+m_2m_1, \nonumber \\
&=&8c^2+d^2-ac \nonumber \\
-S_3&\equiv& m_1m_2m_3 \nonumber \\
&=&ad^2.
\end{eqnarray}
Eliminating $d$ and $c$, the resulting
cubic equation for the parameter $a$ can be written as
\begin{equation} \label{Eq:cubic-a}
9a^3-17S_1a^2+(8S_1^2+S_2)a-S_3=0.
\end{equation}
A physical solution requires $a$ real and positive.
This is equivalent to having three real solutions for $a$.
One of these is negative and two are positive.
At any point inside the allowed domain in the $(m_1/m_3)$--$(m_2/m_3)$
plane (See Fig.~1), there are thus two allowed solutions, 
denoted Solutions~1 and 2.

This diagonalization has to be carried out for both neutrinos and charged
leptons, in order to obtain the neutrino mixing matrix \cite{Gastmans:2002wm}
\begin{equation} \label{Eq:U-matrix}
U=\left[
\begin{array}{ccc}
U_{e1} & U_{e2} & U_{e3} \\
U_{\mu 1} & U_{\mu 2} & U_{\mu 3} \\
U_{\tau 1} & U_{\tau 2} & U_{\tau 3}
\end{array}\right],
\end{equation}
where ($\epsilon=1$ or $i$, depending on whether or not there is CP violation)
\begin{equation}  \label{Eq:V_CKM_lepton}
U=(V_{\rm CKM}^{\ell})^\dagger
=R^{\rm T}(\ell)\,\diag(\epsilon,1,1)R(\nu)
\end{equation}
relates the neutrino mass eigenstates to the flavor states of charged-current
interactions:
\begin{equation}
|\nu_e\rangle=U_{e1}|\nu_1\rangle+U_{e2}|\nu_2\rangle+U_{e3}|\nu_3\rangle,
\quad{\rm etc.}
\end{equation}

There are four possible solutions to Eq.~(\ref{Eq:cubic-a}), two for the
neutrino sector, and two for the charged lepton sector.  Furthermore, the
model has two signs (denoted ``parities'') to be specified: $b$ and $d$ can
each be either positive or negative (unless $d$ is imaginary
\cite{Gastmans:2002wm}).  However, only the product of the ``$b$ parities'' in
the neutrino and charged lepton sector matters, and similarly for the ``$d$
parities'', so we may put both ``parities'' of the charged lepton sector to
$+1$.
\section{Fits to data}

Let us consider first the atmospheric neutrino data.
The Super-Kamiokande results \cite{Fukuda:1998tw} give
$\Delta m^2 \simeq (2-3)\times 10^{-3}$~eV, with
$\sin^2(2\theta)\simeq 1$.
The survival of muon neutrinos is given by
\begin{eqnarray}
\label{Eq:atm-survival}
P_{\nu_\mu\rightarrow \nu_\mu}(t)
&=& 1
-4\biggl[U_{\mu1}^2U_{\mu2}^2\sin^2\left(\frac{\Delta m^2_{21}t}{4p}\right)
+U^2_{\mu1}U^2_{\mu3}\sin^2\left(\frac{\Delta m^2_{31}t}{4p}\right)
\nonumber \\
&&\quad\quad
+U_{\mu2}^2U_{\mu3}^2\sin^2\left(\frac{\Delta m^2_{32}t}{4p}\right)
\biggr] ,
\end{eqnarray}
where $U$ is the neutrino mixing matrix of Eqs.~(\ref{Eq:U-matrix}) 
and (\ref{Eq:V_CKM_lepton}).
In the limit of $\Delta m^2_{21}t/4p \ll 1$ this simplifies,
and invoking further unitarity, one finds
\begin{equation}
P_{\nu_\mu\rightarrow \nu_\tau}(t)
\simeq 4U^2_{\mu3}U^2_{\tau3}\,
\sin^2\left(\frac{\Delta m^2_{32}t}{4p}\right) ,
\end{equation}
which suggests that one needs
$|U_{\mu3}U_{\tau3}|={\cal O}(1)$.
This can be achieved within the model (for both Solutions 1 and 2), 
for $m_1\ll m_3$, with also $m_2$ small compared with $m_3$.
Furthermore, the data suggest that the scale $m_3$ must be 
such that $m_3^2\simeq (2-3)\times 10^{-3}$~eV,
or $m_3={\cal O}(0.05~{\rm eV})$.

Fits to atmospheric data confirm this qualitative analysis.
Invoking also the solar Cl \cite{Cleveland:1994er}, 
Ga \cite{Abdurashitov:1999bv,Hampel:1998xg},
Super-Kamiokande \cite{Fukuda:1998fd,Fukuda:2001nj}
and SNO \cite{Ahmad:2001an,Ahmad:2002jz} neutrino data,
one finds that Solution~2 for the neutrino sector,
Solution~1 for the charged lepton sector
and ``$b$ parity''=$-1$ give good fits for
$m_1\ll m_3$, with $m_2$ also small as compared with $m_3$.
A $\chi^2$ determined from these different atmospheric and solar
survival probabilities leads to good fits (see Fig.~1)
with $m_3$ about 0.052~eV, $m_2$ about 0.01~eV,
and $m_1\sim$ 0.002--0.004~eV
\cite{Osland:2000bh,Osland:2001kp,Gastmans:2002wm}.
The ``$d$ parity'' is unimportant.

\refstepcounter{figure}
\label{Fig:lvls-052}
\addtocounter{figure}{-1}
\begin{center}
\setlength{\unitlength}{1cm}
\begin{picture}(15.6,6)
\put(-0.5,0.0){
\mbox{\epsfysize=6.5cm\epsffile{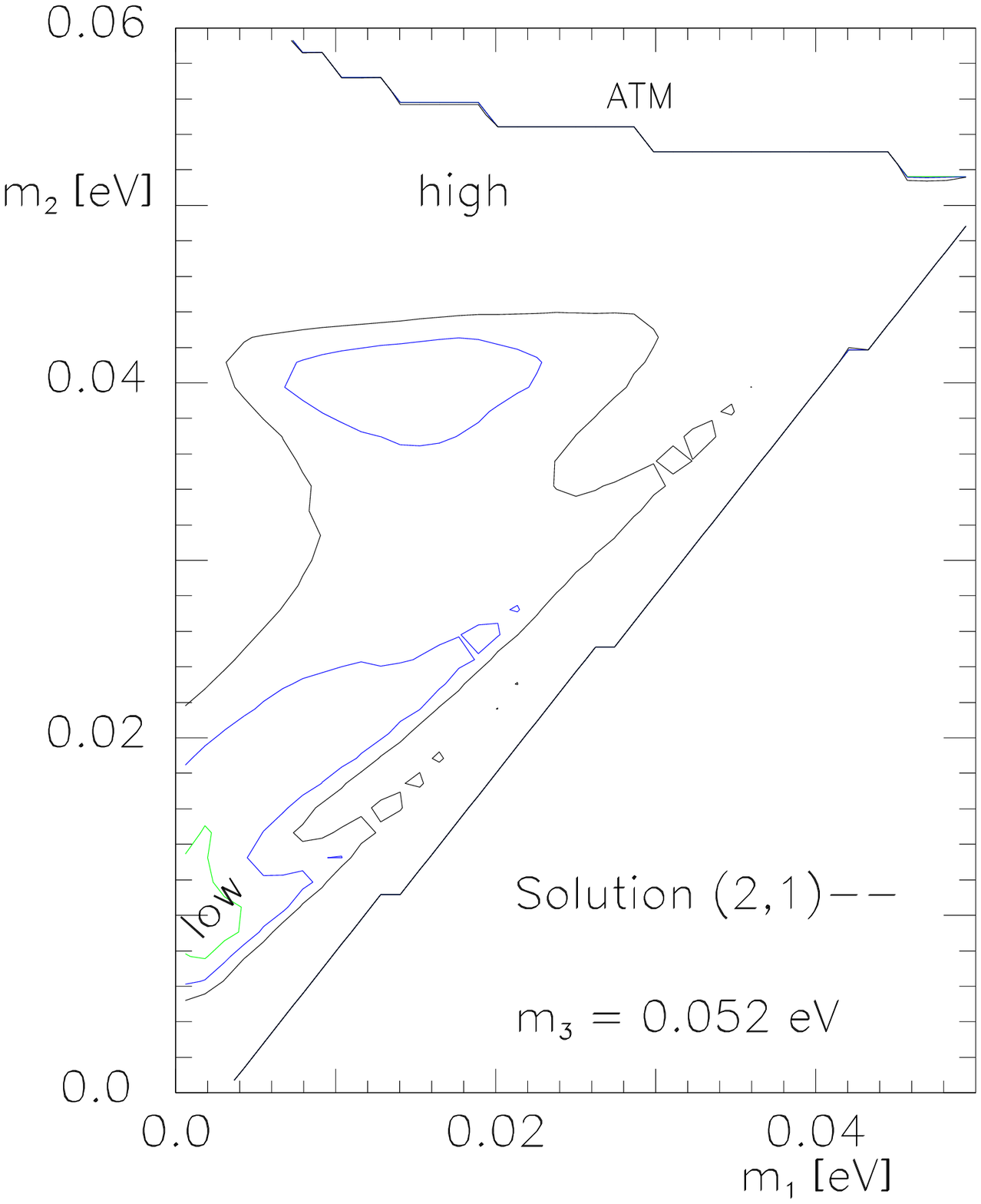}}
\mbox{\epsfysize=6.5cm\epsffile{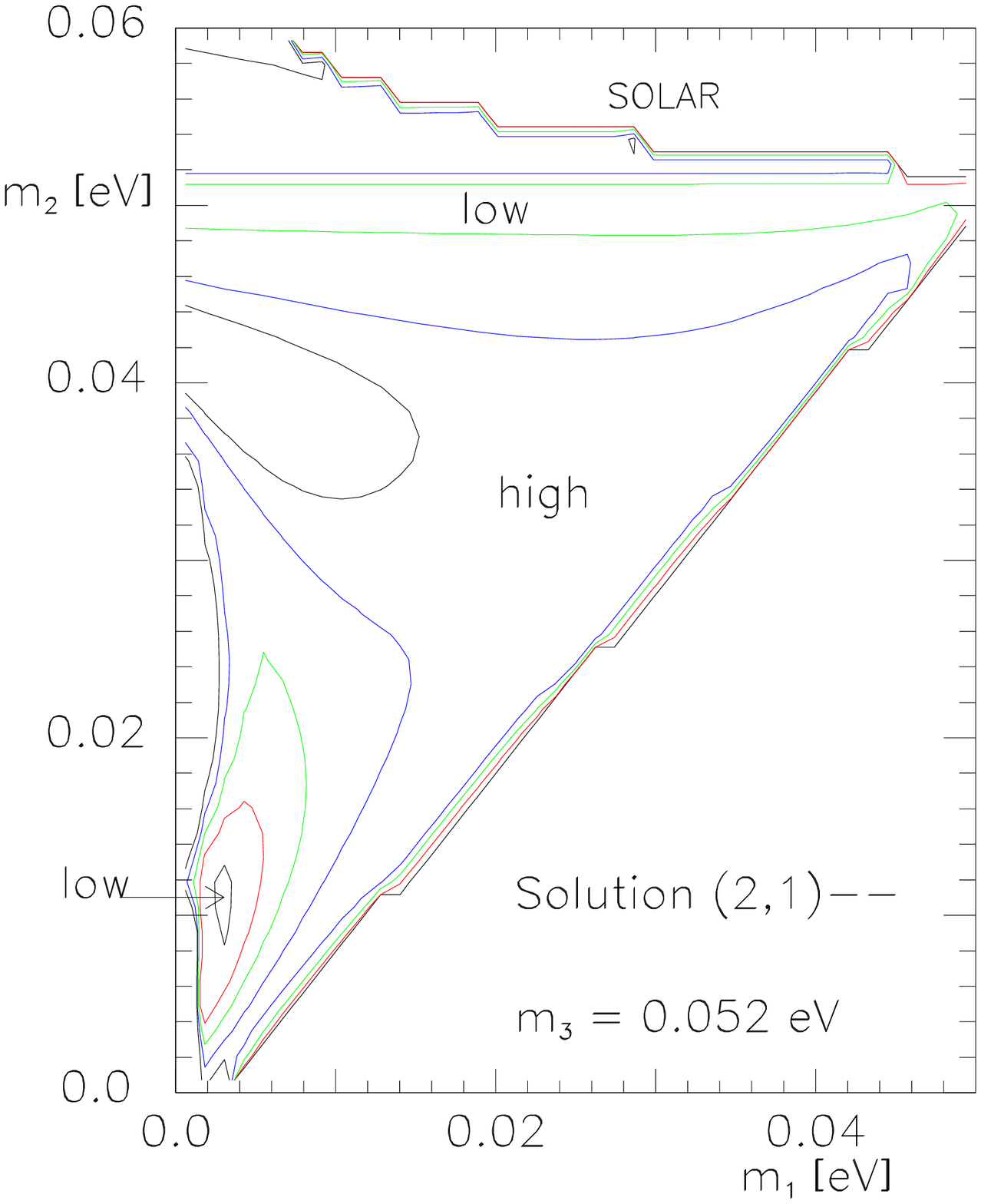}}
\mbox{\epsfysize=6.5cm\epsffile{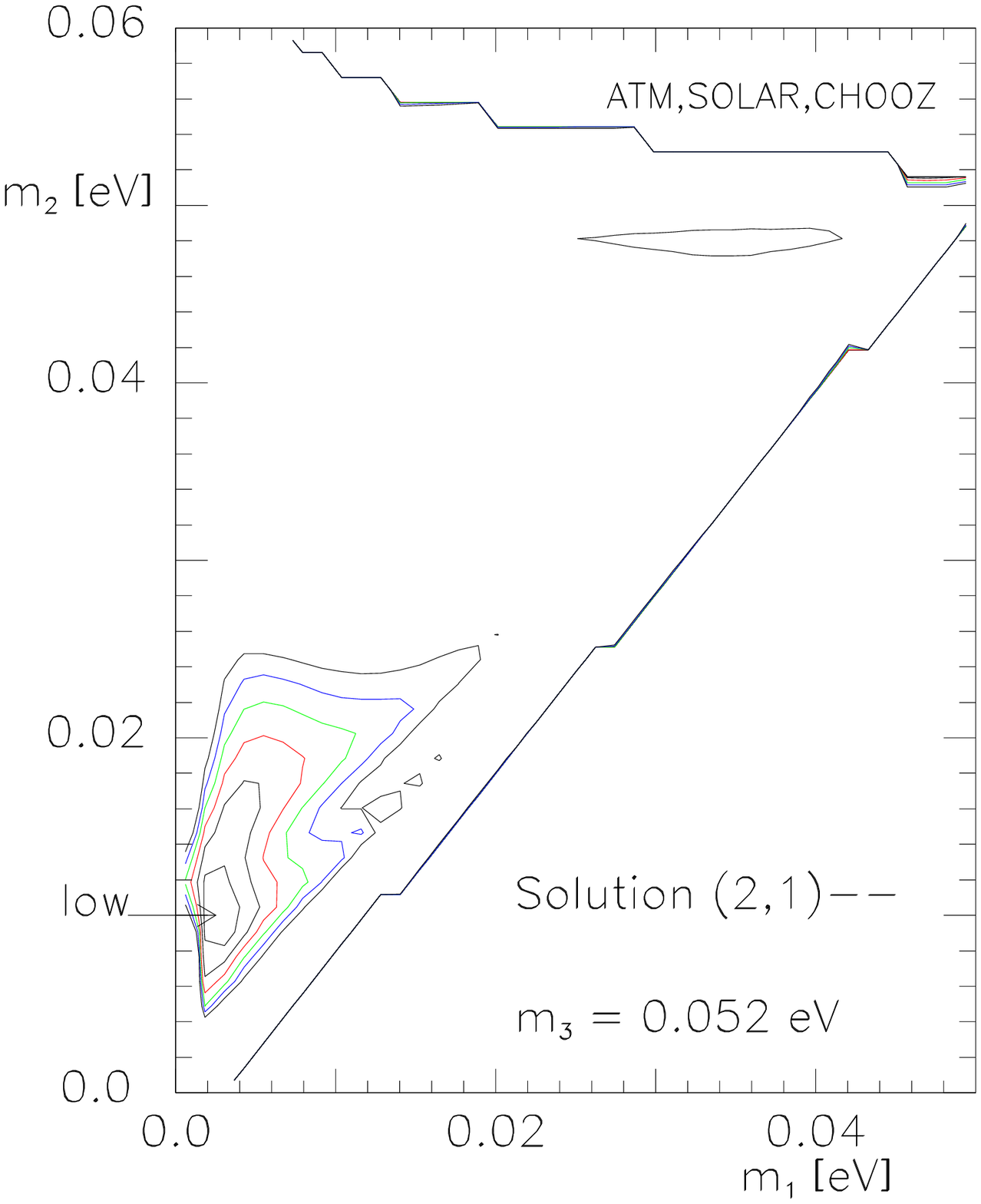}}}
\end{picture}
\end{center}
\vspace*{2mm} {\small {\bf Figure~1.} Fits to the atmospheric and solar
neutrino data for Solution 2 (neutrino sector) and 1 (charged leptons) and
both ``parities'' negative.  Contours are given at $\chi^2=5$, 10, 15, 20, 25.
Sum (right panel) also at 30, 35, 40, 45, 50. The best-fit point (right
panel, at the lower left), is marked ``low''.}
\vspace*{4mm}

The effect of the CHOOZ data \cite{Apollonio:1997xe}
is to disfavor the region $m_2={\cal O}(m_3)$, where $|U_{e3}|$
is large. But this region is already disfavored by the solar 
neutrino data, as is seen in the middle panel of Fig.~1.
In terms of the more conventional two-flavour analyses for 
the solar-neutrino sector, these fits roughly correspond to 
the large-mixing-angle solution.
\section{Summary}
Analytic results for the solutions to the MSW equations for three neutrino
flavours are very valuable for a fast scanning over the parameters of some
given model for the mass matrix.

The LNW mass matrix is a very constrained model that in the quark sector
describes the CKM matrix, and in the neutrino sector gives the mixing in terms
of the mass eigenvalues. Applied to the neutrino data, the model gives
a very good fit.

The solar neutrino data has also been studied within the same model, 
using numerical integration methods (no ${}_2F_2$'s) \cite{Osland:2000gi}.
An additional fit was then found at $m_1\simeq 2.8\times10^{-6}$~eV, 
corresponding to the small-mixing-angle solutions.
However, this point is disfavoured by the atmospheric neutrino data,
and by the flat electron recoil spectrum \cite{Fukuda:1998ua}.
\bigskip

\leftline{\bf Acknowledgments} \par\noindent It is a pleasure to thank the
organizers of ``Beyond the Desert 2002'' for creating a very stimulating
atmosphere.  I should also like to thank R. Gastmans, G. Vigdel and T. T. Wu
for very fruitful collaborations.  This work is supported in part by the
Research Council of Norway.
\section*{References}

\end{document}